\documentclass[journal]{IEEEtran} 

\usepackage{graphicx}
\usepackage{amsmath}
\usepackage{tabu}

\begin{document}

\title{Measuring Water Vapor and Ash in Volcanic Eruptions with a Millimeter-Wave Radar/Imager}

\author{Sean~Bryan$^*$\thanks{$^*$Email: sean.a.bryan@asu.edu}, Amanda~Clarke, Lo{\"y}c~Vanderkluysen, Christopher~Groppi, Scott~Paine, Daniel~W.~Bliss,~\textit{Fellow,~IEEE}, James~Aberle,~\textit{Senior~Member,~IEEE}, and Philip~Mauskopf}

\maketitle

\begin{abstract}
Millimeter-wave remote sensing technology can significantly improve measurements of volcanic eruptions, yielding new insights into eruption processes and improving forecasts of drifting volcanic ash for aviation safety. Radiometers can measure water vapor density and temperature inside eruption clouds, improving on existing measurements with infrared cameras that are limited to measuring the outer cloud surface. Millimeter-wave radar can measure the 3D mass distribution of volcanic ash inside eruption plumes and their nearby drifting ash clouds. Millimeter wavelengths are better matched to typical ash particle sizes, offering better sensitivity than longer wavelength existing weather radar measurements, as well as the unique ability to directly measure ash particle size in-situ. Here we present sensitivity calculations in the context of developing the WAMS (Water and Ash Millimeter-wave Spectrometer) instrument. WAMS, a radar/radiometer system designed to use off-the-shelf components, would be able to measure water vapor and ash throughout an entire eruption cloud, a unique capability.
\end{abstract}

\section{Introduction}

Observations of volcanic ash and gas provide key data needed for the study of volcanic eruptions. In addition to providing insights into geophysical processes at work in volcanoes, the disruption and hazard to worldwide aviation caused by the 2010 Eyjafjallaj{\"o}kull eruption in Iceland has brought to the forefront the practical need for improved real-time data and modeling for volcanic eruptions \cite{bonadonna11}. For both fundamental research and aviation safety, there are several numerical codes, such as ATHAM \cite{textor06}, that model the 3D content and dynamics of both the main volcanic eruption plume, and the high-altitude drifting ash cloud. There are also codes, such as Ash3d \cite{mastin13}, that calculate the distribution of ash after large eruptions at global distance scales relevant for aviation safety. As inputs to simulate specific eruptions, and to validate the model accuracy, many of these codes rely on measurements of the temperature, ash mass distribution, gas temperature, water vapor and gas content of eruptions at the vent. Since an ash cloud can spread globally, potentially disrupting aviation, direct high-fidelity measurements of ash mass distribution and particle size near the eruption would significantly improve ash distribution forecasts for aviation. Measurements of the cloud would also improve fundamental understanding of eruption processes \cite{vaneaton15}.

Currently available measurement techniques have limitations in their ability to quantitatively inform these eruption models. Optical \cite{sparks82} and infrared \cite{harris13} imaging techniques offer good spatial resolution and give an indication of the temperature distribution of material in the high-altitude cloud. However, due to the cloud's opacity in the optical and infrared, these methods are only able to measure the outer cloud surface, and not the interior. Conventional weather radar technology \cite{schneider13,corradini16,maki16,marzano13,arason11,donnadieu05} uses centimeter-wavelength radio, which means it is best suited to measure larger ash particles at higher concentrations that typically are in the main eruption column near the vent. However, conventional weather radar is much less sensitive to fine ash, which may reach higher elevations in the eruption column and drifting cloud. Lidar has been used to measure fine ash in volcanic eruptions \cite{scollo12,scollo15}. Interpreting these reflectivity measurements in terms of ash mass relies on theoretical modeling and some assumptions, since their operating wavelengths are very different from the typical sizes of volcanic ash particles. This means that both lidar and conventional weather radar can not directly measure the ash particle size.

Millimeter-waves are uniquely suited to improve volcanic eruption measurements. In this paper, we discuss how millimeter-wave radiometers could be used to image the water vapor density and temperature distribution, and how millimeter-wave radar could be used to map the ash mass density and particle size. These sensitivity calculations are motivated by our development of the WAMS (Water and Ash Millimeter-wave Spectrometer) instrument concept.

\begin{figure*}
\begin{center}
\includegraphics[width=0.58\textwidth]{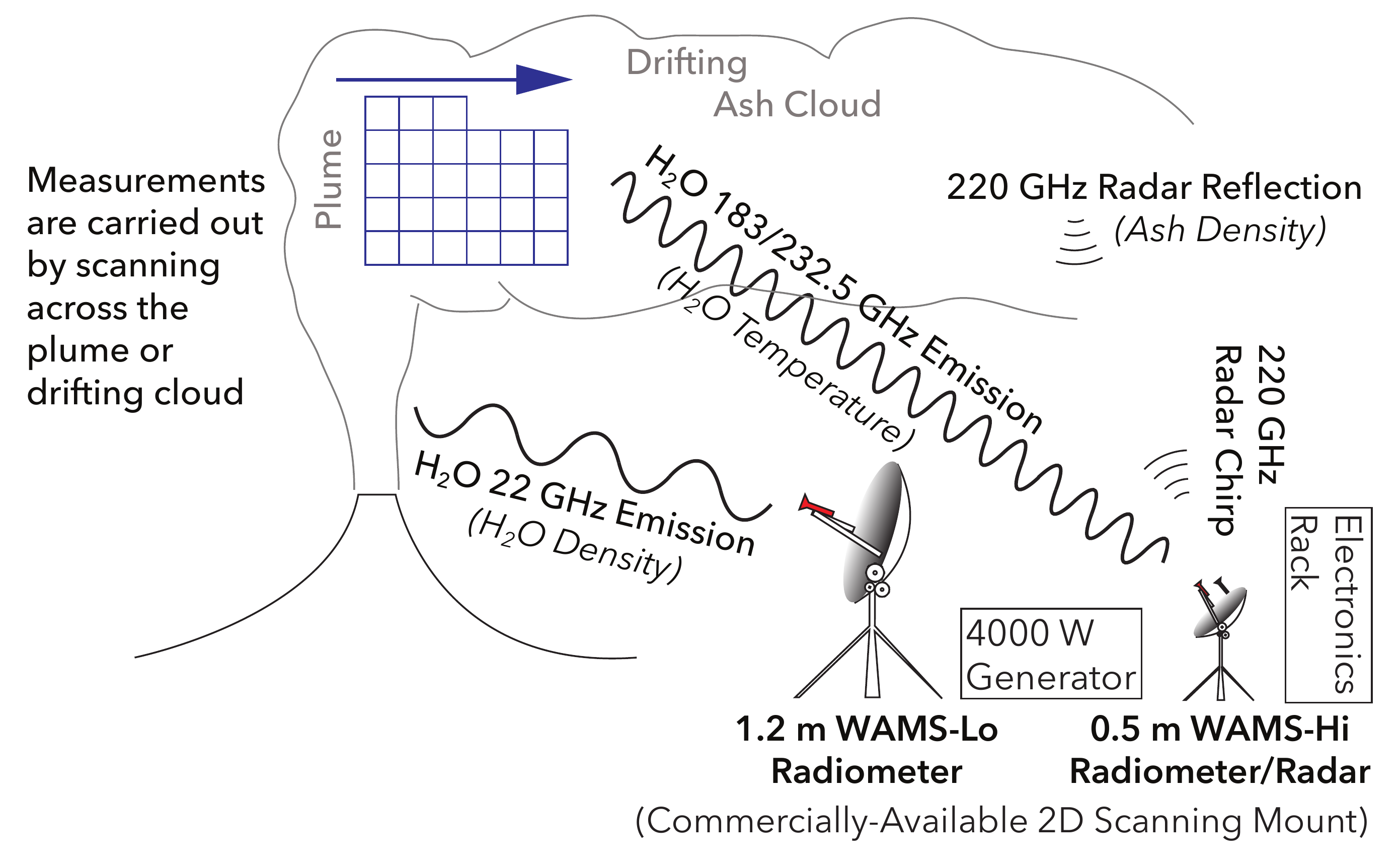}
\caption{Illustration of the WAMS instrument concept. Millimeter-wave radar will enable 3D imaging of the fine ash of high-altitude volcanic eruption clouds for the first time. Microwave radiometers at several frequencies will measure the density and temperature distribution of water vapor throughout the eruption. The instrument would be compact enough to be rapidly transported and deployed near an eruption as it happens. \label{wams_overview}}
\end{center}
\end{figure*}

The proposed WAMS instrument will contain both radar and radiometer systems. The radiometers will consist of a low-frequency imager with 8 bands at 18-26 GHz, and a high-frequency 180-235 GHz spectroscopic imager. These two imaging systems will look inside the volcanic eruption, rapidly scanning to map the interior density and temperature distributions of the water vapor. This will be accomplished by simultaneously monitoring three different emission lines of water (22 GHz, 183 GHz, and 232.5 GHz). The measured relative intensity of the three lines indicates the vapor temperature, and absolute strength of the 22 GHz line indicates the density.

WAMS will also have a 220 GHz radar system. The radar operating frequency was selected to maximize the returned signal from typical ash particle sizes in the main eruption plume, and also to improve the sensitivity to drifting fine ash clouds. Realtime retuning of the radar frequency (discussed in Section \ref{ash_mass}) will enable in-situ measurement of particle size at different points in the eruption plume and drifting cloud, a measurement that is not possible with existing methods.

The WAMS instrument will be portable, with all the equipment fitting in a standard pickup truck for transportation to an observing site near the erupting volcano. Power will be from a standard gas generator. The low frequency imaging system will be a 1.2 m diameter dish antenna installed on a tracking mount from ORBIT Communications, followed by a B\&Z 22 GHz low noise amplifier, bandpass filters, and diode power detectors to measure the signal. The high frequency imaging/radar system will be a 0.5 m diameter dish antenna installed on a precision tracking mount from RIE Technologies. The radar transmitter will be a 220 GHz amplifier multiplier chain (AMC) from Virginia Diodes, connected to one polarization of an orthomode transducer (OMT) \cite{reck13,navarrini10} and feed horn \cite{tan2011a, tan2011b} we will fabricate at Arizona State University. A 220 GHz Virginia Diodes WR5.1 MixAMC receiver will be attached to the other polarization of the OMT. The OMT will isolate the receiver from the transmitter, and a Faraday rotator system from QMC will couple the polarized reflection from the target onto the orthogonally-oriented polarization sensitivity of the receiver. When operating as a radar, the RF output of a CASPER/ROACH board will drive the RF input port of the transmitter with a Frequency Modulated Continuous Wave (FMCW) radar chirp. The RF input of the board will then sample the radar returns received from the target. When operating as a radiometer, only the RF input side will be used. Except for the OMT and feed horn which some of the authors \cite{navarrini10,tan2011a,tan2011b} have experience successfully fabricating, all of the other components of the WAMS system are available commercially. A conceptual overview of WAMS is shown in Figure~\ref{wams_overview}.

\section{Water Vapor in Volcanic Eruptions}

\subsection{Scientific Impact}

Water vapor is typically the most abundant gas released during volcanic eruptions and is the primary driver of magma fragmentation and particle acceleration at the vent. Measurements of water vapor probe degassing processes that initiate explosive eruptions. Also, given the fact that the solubilities of SO$_2$ and H$_2$O in magmas are fundamentally different (meaning existing SO$_2$ measurements are an incomplete proxy for understanding degassing processes that control large-scale dynamics), water vapor measurements can help to complete our picture of volcanic degassing processes immediately prior to, during and after explosive eruptions \cite{kazahaya94,steffke11}.

Condensation of water vapor also plays a critical role in ash aggregation. Imaging water vapor temperature will trace temperature variation within plumes, and along with the water vapor concentration measurement, help detect zones in which condensation and therefore ash aggregation may be important. This data would help constrain the heat budget of volcanic plumes, in turn improving understanding of air entrainment and subsequent buoyancy development. In particular, we envision that water vapor concentration measurements in the main plume will provide good estimates of magmatic water content, especially since the instrument should be able to measure water vapor properties near the vent itself. Comparison between cross-section measurements at different heights will provide a measure of plume dilution, thus constraining the entrainment of ambient atmosphere.  A combination of water vapor concentration and temperature higher in the rising plume will provide clues regarding the location of intense condensation zones where particle aggregation is expected to occur \cite{woods93,glaze97,herzog98,textor06}.

\subsection{Radiometer Measurements}
\label{wams_radiometer}

\begin{figure*}
\begin{center}
\includegraphics[width=0.73\textwidth]{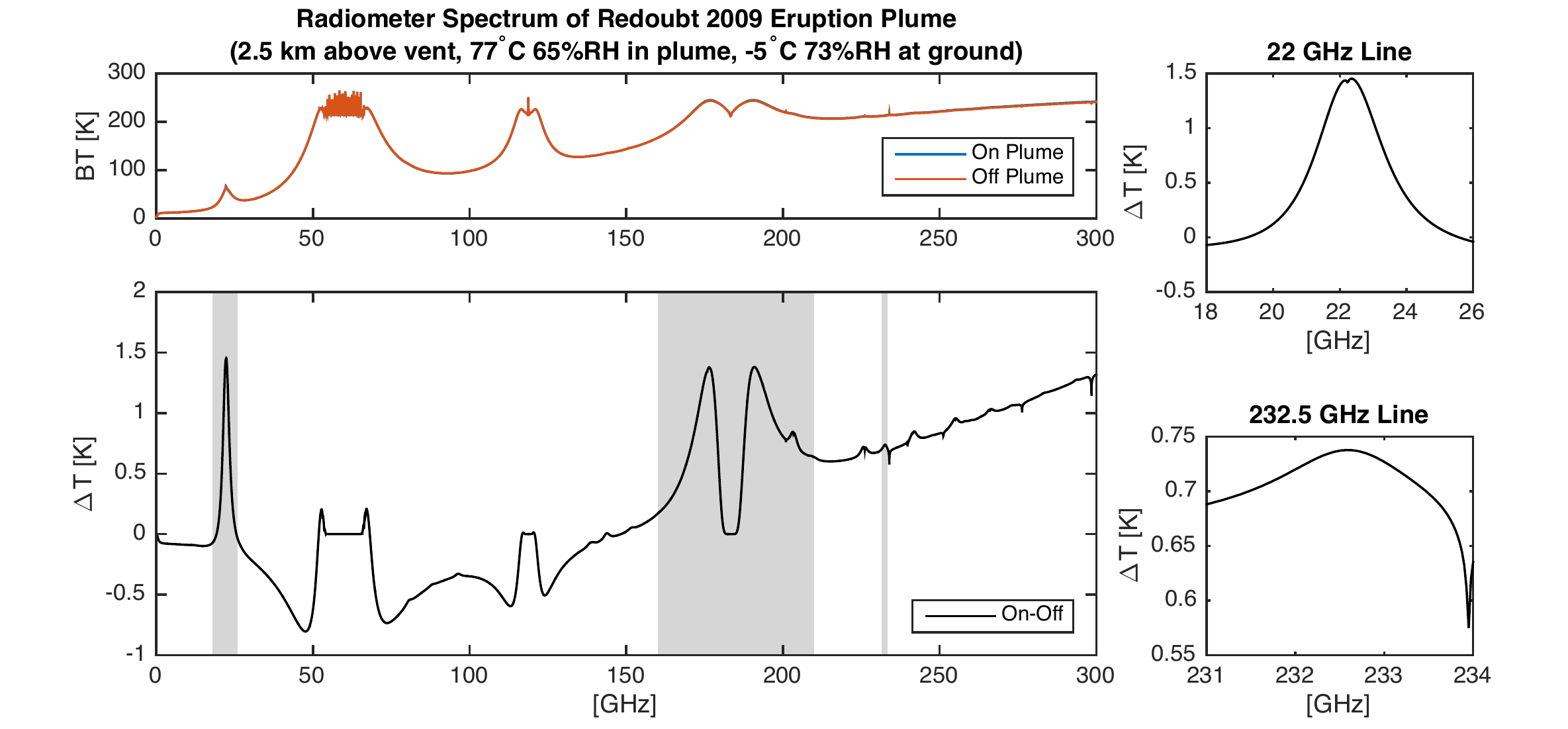}
\caption{Simulated radiometer spectrum from 10 km standoff distance observing the water vapor in the Redoubt 2009 eruption. The warm humid eruption (77$^\circ$C and 65\% RH at this point) creates a contrast in millimeter-wave water lines against the cool dry foreground and background (-5$^\circ$C and 73\% RH at the ground, a similar relative humidity, but lower absolute amount of water vapor). An imaging spectrometer, such as WAMS, would take data like this at all altitudes above the vent as well as in cross sections across the eruption. The top-left panel shows the Brightness Temperature (BT) spectrum with the instrument pointed 2.5 km above the eruption vent. The 22 GHz, 183 GHz, and 232.5 GHz H$_2$O lines are highlighted in grey bands in the difference spectrum (bottom-left) between the spectra taken on and off the eruption plume (top-left). (The sharp feature at 233.95 GHz is due to atmospheric O$^{18}$O.) The two right panels show that the 22 GHz and 232.5 GHz lines would be easily detectable (instrument noise discussed in Section \ref{wams_radiometer}) by a purpose-built millimeter-wave radiometer with off-the-shelf components, as would the large signal in the wings of the strong 183 GHz line. \label{radiometer_redoubt}}
\end{center}
\end{figure*}

A millimeter-wave radiometer instrument would be well suited to addressing these questions. Measuring water vapor line emission is a conceptually similar technique to Differential Optical Absorption Spectroscopy (DOAS), a well-established technique used in atmospheric and geophysical remote sensing \cite{edner94, weibring98, galle99}. The millimeter-wave approach is still a differential path measurement, but with the spectral lines of the target molecules in emission rather than absorption. A promising target line is the 22 GHz (i.e. 13.6 mm wavelength) emission line of water. This passive measurement method would quantify the column density (sometimes referred to as path-integrated concentration) of water vapor in the antenna's main beam. The relatively weak 22 GHz water line is optically thin (i.e., semi-transparent) even for paths that pass completely through the Earth’s atmosphere \cite{kursinski12}, meaning that foreground atmospheric attenuation will not limit the measurement range. The strength of the 22 GHz spectral feature is proportional to the column density of water molecules, but depends little on the gas temperature. A differential measurement can be performed to measure the difference in water vapor column density between a line of sight through a volcanic eruption and a reference line of sight away from the eruption. This method has a distinct advantage over open path Fourier Transform Infrared (OP-FTIR) spectrometry, because it does not require the presence of an active infrared source behind the plume.  

Other water lines can be used to measure the temperature of water vapor in the eruption. The amplitude of the 183 GHz and 232.5 GHz (1.64 mm and 1.29 mm wavelength) water emission lines (see Figure \ref{radiometer_redoubt}) are temperature-dependent. The 183 GHz line is optically thick at line center, but the emission temperature can be probed by measuring frequencies in the wings of the line where foreground atmospheric attenuation is low. By contrast, the 232.5 GHz “hot” water line is optically thin, and located at a frequency in the $\sim$190-280 GHz atmospheric window (i.e. low atmospheric foreground attenuation). This line has a high excitation temperature (3450 K), so it probes the plume temperature via its strength relative to that of the 22 GHz line. A millimeter-wave radiometer covering these three lines could make the first direct measurements of temperature inside a volcanic eruption, as existing thermal imagers cannot penetrate due to their much shorter operating wavelengths (typically 7-16 $\mu$m). 

We built a simulation pipeline to calculate what millimeter-wave spectra would look like when observed with a radiometer. The main eruption plume, to first-order, can be approximated as a steady, axi-symmetric plume in which particles and gas are in thermal and mechanical equilibrium. As such, in our current pipeline, plume diameter, ash and water vapor concentration, mixture temperature, and other properties are calculated as a function of altitude using the Plumeria code \cite{mastin07}. To simulate the 2009 eruption in Redoubt, Alaska, we used a default scenario file included with the Plumeria code.

The resulting atmospheric conditions, and plume water vapor density and temperature distributions, were input into \textit{am} \cite{paine14} to simulate the radiometer signal. The \textit{am} code can calculate the attenuation and thermal emission of arbitrary layers of atmospheric and other gasses for radio and millimeter wavelengths. Since any thermal emission by volcanic ash would be small, broadband, and would not contribute to the strengths of millimeter wave lines, for this study we do not consider the thermal emission of the ash. As a concrete example, to generate Figure \ref{radiometer_redoubt} we considered a standoff distance of 10 km and calculated the signal that WAMS would see if it was pointed up along a line of sight passing 2.5 km above the vent. The radiometer therefore would be looking up through foreground atmosphere, with the temperature and pressure changing as the altitude increased. Then, the line of sight would pass up through the eruption plume itself. Finally, the line of sight would continue ascending up through the background atmosphere behind the eruption. This was modeled in \textit{am} with 20 layers of foreground air following the US Standard Atmosphere Model (scaled to the weather conditions at the site during the eruption) 10 layers inside the eruption with data from Plumeria, and 20 more layers of background atmosphere. This approach means that foreground attenuation, as well as both foreground and background water vapor emission, are integrated into the simulation. Figure \ref{radiometer_redoubt} shows the calculated radiometer signal pointed at the plume, the background signal when the instrument is pointed off the plume (i.e. replacing the 10 model layers inside the eruption, with 10 layers of US Standard Atmosphere), and the difference between the two spectra. Since the beam is small when projected at this distance, we do not convolve this simulation with the instrument beam. It varies with frequency, but as discussed in Figure~\ref{radar_iceland} it is 60 m at 220 GHz at 10 km range. This differencing method is a standard approach in both astronomical spectroscopy and in DOAS measurements in volcanology and remote sensing.

We will take data on and off the plume at a wide range of elevations above the vent. Our simulation pipeline already indicates how this large rich dataset might look: it would be a new Figure~\ref{radiometer_redoubt} at each elevation and horizontal point in the eruption. This full simulation yielded the insight that while the 22 GHz line will trace vapor density in the expected linear way, extracting the vapor temperature from the measured strengths and widths of the 22, 183, and 232.5 GHz lines will require a fitting code to be developed to handle the foregrounds. This fitting code will utilize the simulation code presented here as a starting point, but it will likely also be driven by field data and is therefore beyond the scope of this paper.

The off-the-shelf components used in WAMS, or a similar instrument, would easily be able to detect these signals. The radiometer equation
\begin{equation}
\mathrm{Noise~Level} = \frac{T_s + T_b}{\sqrt{B \tau}},
\end{equation}
yields the noise level of a radiometer with a system temperature $T_s$, bandwidth $B$, and integration time $\tau$, when observing a scene with a background temperature $T_b$. Gain fluctuations inside a receiver can increase the noise level above this fundamental limit, so we will investigate using standard modulation techniques to remove the effect of these fluctuations if the intrinsic stability of the WAMS receivers does not achieve this performance.

With a system temperature of 130 K (data sheet from the B\&Z low noise amplifier) and a channel bandwidth of 100~MHz, conservatively assuming a 300 K background temperature the WAMS 22 GHz radiometer would reach a noise level of $43$~mK in a 1~s observation. For the 183/232.5~GHz radiometer system, a system temperature of 2500 K (measured in the lab) would have a noise level of $280$~mK in 1~s. We will use 50~MHz bandwidth channels, but since the higher frequency band in WAMS is a dual-sideband heterodyne receiver, the actual noise bandwidth is doubled. This means that even when observing 10 km away from the plume, and pointing 2.5 km in elevation above the vent (as illustrated in Figure~\ref{radiometer_redoubt}), the 43-280 mK noise level of the WAMS radiometers would clearly detect the $\sim1$ K water vapor lines, as well as stronger features from higher density/temperature points in the eruption than the one we considered in Figure \ref{radiometer_redoubt}. In cases where the lines are weaker (e.g., the 232.5 GHz line in cooler plumes) the integration time per point could be increased to compensate, at the penalty of reduced ability to image eruption dynamics. Along with simultaneous IR, optical, and SO$_2$ measurements with existing instruments, imaging volcanic eruption clouds in these three millimeter-wave lines, at high signal-to-noise, with a radiometer will open an important new window on the interior dynamics of eruptions.

\section{Volcanic Ash}

\subsection{Scientific and Aviation Impact}

In 1995, Volcanic Ash Advisory Centers (VAACs) were established to track and predict transport and dispersal of volcanic particles (pulverized magma) to ensure safety in aviation. Volcanic particles are so damaging to aircraft engines that all engine and aircraft manufacturers have zero tolerance policies with regard to exposure to volcanic ash clouds. VAACs use a variety of Volcanic Ash Transport and Dispersion (VATD) models to forecast where and if volcanic ash will be a hazard to aviation. There are several measurements, not currently available, that would improve these forecasts and improve fundamental understanding of eruption processes. Quantification of ash concentration near an eruption vent, along with independent column ascent velocity measurements by profile comparison or other methods, would provide an estimate of mass flux to serve as input for VAAC forecasters. Comparison of ash concentration at different heights and times within the rising plume and drifting cloud will provide excellent datasets for comparison against numerical codes designed to simulate explosive eruptions. These comparisons will yield information about the combined effects of ambient air entrainment plus sedimentation. The high altitude drifting ash cloud is the most critical eruption zone for understanding the long-range hazards to commercial aircraft. This fine-grained dilute drifting cloud, although critically dangerous to aircraft, is typically invisible to existing weather radar instruments. Also, as discussed in Section \ref{particle_size_text}, existing weather radar is not able to directly measure the ash particle size distribution.

\begin{figure*}
\begin{center}
\includegraphics[width=0.82\textwidth]{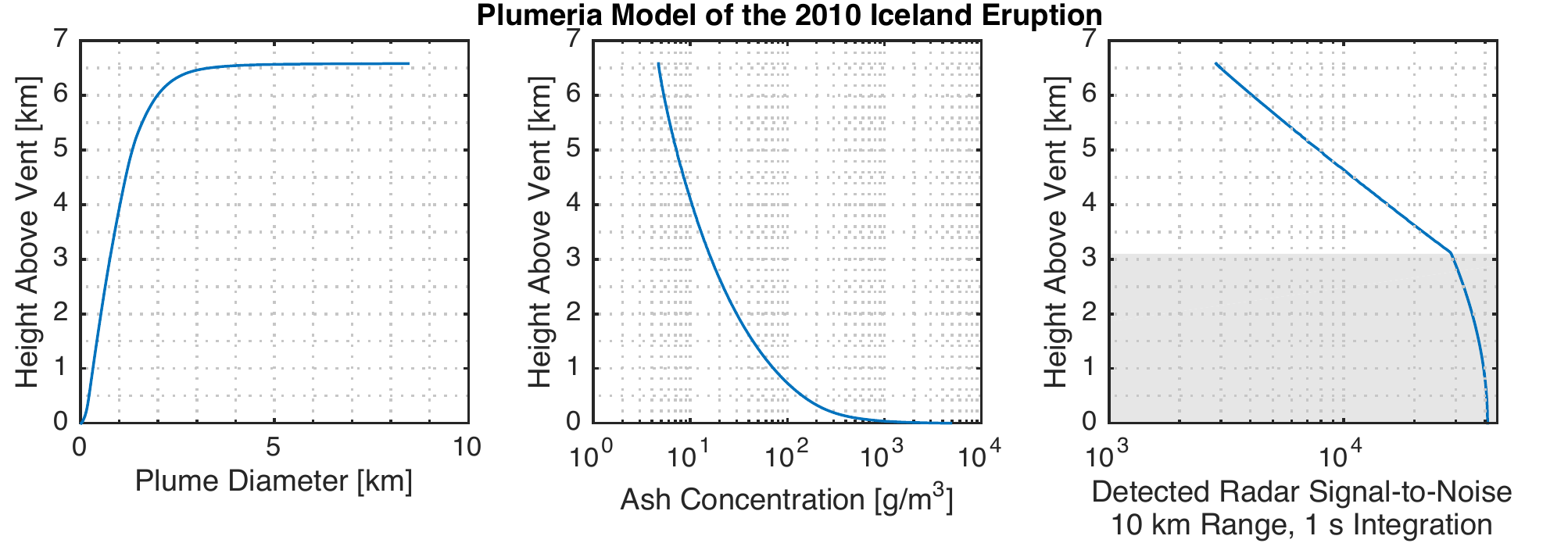}
\caption{Simulated radar measurements of volcanic ash if WAMS had observed the 2010 Eyjafjallaj{\"o}kul eruption in Iceland. Plumeria was used to calculate the main eruption plume diameter (left panel) and ash density (center panel) profiles above the volcano vent. The radar cross section was calculated with a Mie scattering code (results shown in Figure~\ref{rcs_particle_size}), and the \textit{am} atmospheric propagation code was used to model atmospheric attenuation (for the conditions at the site at the time, the attenuation would have been 1.22 dB/km). Integrating these three codes yielded the calculated WAMS radar signal (right panel) from volcanic ash in a cubic volume element observed from 10~km away (40~m on each side, similar to the $\sim60$~m WAMS beam size at 10~km range). The grey region highlights where WAMS will only receive radar returns from the outer surface of the eruption because the ash density is too high for the millimeter-waves to penetrate further. Existing weather radar can measure this lower part of the plume, and the WAMS measurements will still have high signal-to-noise even up to the top of the main plume and in the drifting fine ash cloud. \label{radar_iceland}}
\end{center}
\end{figure*}

\subsection{Radar Signals from Ash}

\subsubsection{Detectability}

Although radar systems have previously been used to measure ash concentration in the atmosphere \cite{harris83, rose95, schneider13}, the weather radars currently used for this purpose work at non-optimal frequencies and spatial resolution for measurement of volcanic eruptions. Radar measurements at 220 GHz (1.36 mm wavelength) would be more closely matched to the typical size of ash particles ($\sim$50 $\mu$m - 2 mm) than conventional 5 GHz ($\sim6$ cm wavelength) weather radar, and are therefore far more sensitive to particulate concentration. As an approximate illustration of how radar signal can scale with operating wavelength $\lambda$, assuming backscattering from a single dielectric sphere with a diameter $D$ and making the Rayleigh approximation ($D \ll \lambda$), the radar cross-section (RCS) is
\begin{equation}\label{rayleigh_rcs}
\mathrm{RCS} = \frac{\pi^5}{\lambda^4} \left| \frac{\epsilon_r - 1}{\epsilon_r + 2} \right|^2 D^6,
\end{equation}
where $\epsilon_r$ is the dielectric constant of the sphere \cite{marzano06}. Because of the $\lambda^{-4}$ scaling of the signal, a 220 GHz operating frequency results in over \textit{6 orders of magnitude increase} in signal for fine (i.e. much smaller than the wavelength) ash particles when compared to conventional 5 GHz weather radar. For all of the quantitative radar cross section calculations in the rest of this paper, we used a full Mie scattering code (discussed in Section \ref{ash_mass}) that is accurate for any wavelength or particle diameter.

This large reflected signal means that millimeter-wave radar can be used to image very low concentrations of fine ash which are invisible to conventional radar systems. As an example, shown in Figure \ref{radar_iceland}, we simulated the 2010 Eyjafjallaj{\"o}kul eruption in Iceland using Plumeria (and the nominal initial conditions file included with the code for that eruption) as it would have appeared to the WAMS radar observing from 10 km away. The 80 mW transmit power (from the transmitter data sheet), 0.5 m dish, and 2500 K noise temperature (measured in the lab) of WAMS will allow it to obtain very high signal-to-noise ($>$1000) all the way to the top of the 6.5 km simulated plume, even at relatively low ash concentrations of $\sim 1$~g/m$^3$. Scaling this result indicates that even the very low concentrations ($\sim$ mg/m$^3$) relevant for aviation safety will be detectable by WAMS.  This illustrates the complementarity of millimeter-wave measurements to existing weather radar techniques. Weather radar can easily measure large ash grains near the base of the eruption column where millimeter-wave radar cannot penetrate. However, existing weather radar is not sensitive enough to measure fine-grained dilute portions of the main plume, or the higher altitude dilute drifting ash clouds \cite{vaneaton15}.

\subsubsection{Converting the Measured Radar Signal to Ash Mass}
\label{ash_mass}

The Rayleigh scattering model in Equation~\ref{rayleigh_rcs} is accurate for modeling the backscattering from particles much smaller than the wavelength. However, since volcanic ash particles can also be a similar size or larger than the 1.4 mm operating wavelength of the radar, we calculated the scattering from volcanic ash to our pipeline using a Mie scattering code \cite{maetzler02} which is accurate for any particle size. While it is possible to calculate scattering from other particle shapes (e.g., see \cite{weinman06} for a study of millimeter-wave scattering from snowflakes, or \cite{hagfors64} for the correction factor for spheres with Gaussian surface roughness) here we approximate the volcanic ash particles as spherical. We assumed that the dielectric constant of the ash was $\epsilon_r=5.4-0.16i$, consistent with existing experimental data at centimeter and millimeter wavelengths \cite{oguchi09,rogers11}.

\begin{figure}
\begin{center}
\hspace{0.08in}\includegraphics[width=0.43\textwidth]{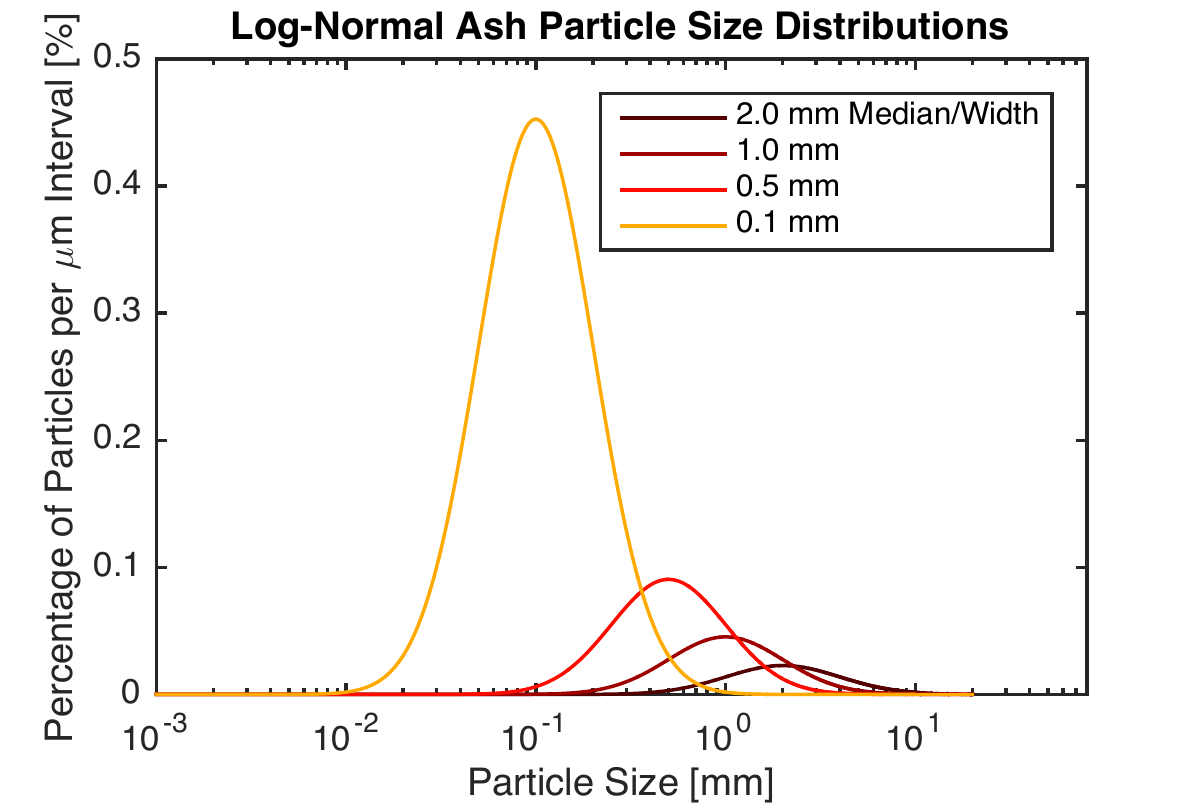}\vspace{0.08in}
\includegraphics[width=0.45\textwidth]{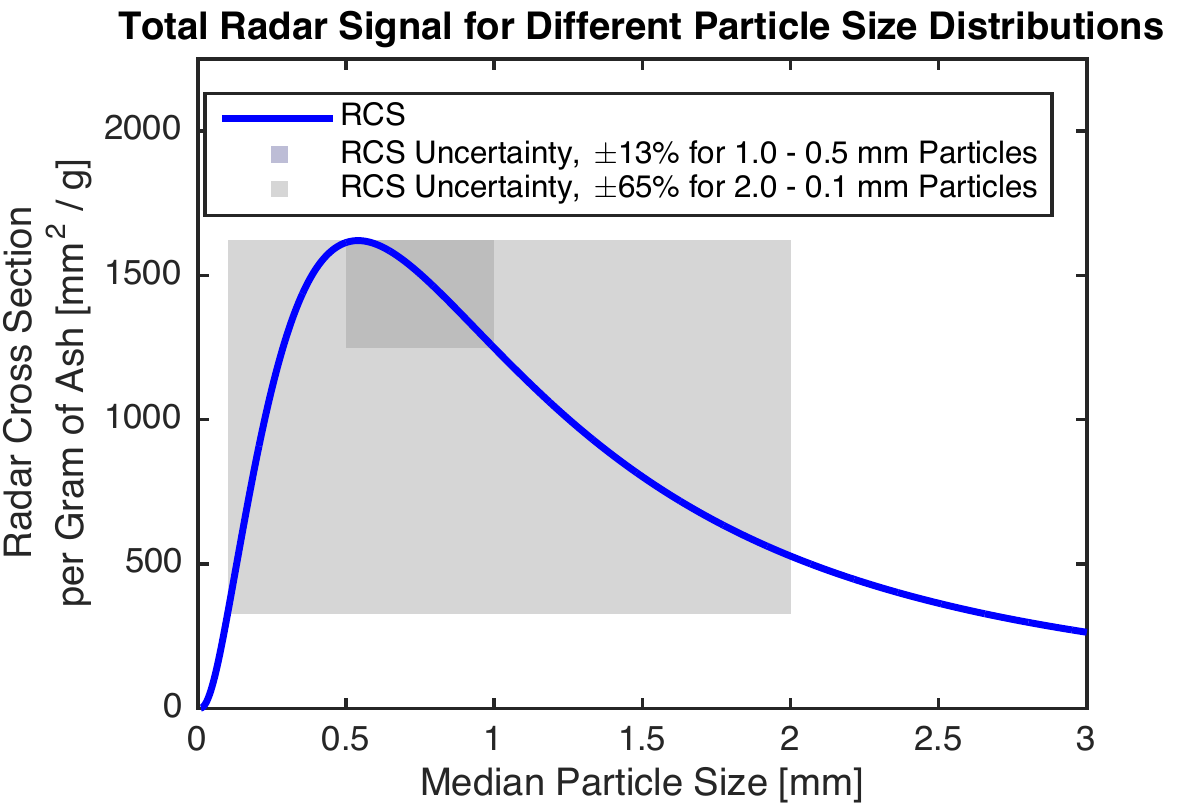}
\caption{Calculated radar signal from volcanic ash, using a Mie scattering code. The top panel shows log-normal particle size distributions, representative of measured volcanic ash particle size distributions \cite{vaneaton15, gudmundsson12}. The median ($\mu$) and width ($\sigma$) parameters of the log-normal distributions are equal, which is consistent with measured distributions. The bottom panel shows the calculated 220 GHz radar signal, per gram of ash, for different particle size distributions (again with the medians and widths equal). This calculated conversion factor will be used to convert measured radar signal (along with measured particle size as illustrated in Figure \ref{rcs_retuning}), into a measurement of ash mass concentration. Even if the particle size is uncertain in over a relatively broad 1.0-0.5 mm range, the uncertainty is only $\pm13\%$ in the conversion from radar signal to mass density. Figure \ref{rcs_retuning} shows that realtime radar retuning can be used to measure the particle size in-situ, reducing this small calibration uncertainty even further. \label{rcs_particle_size}}
\end{center}
\end{figure}

Ignoring resonance effects (which can be significant, and are treated in our Mie scattering code), two limiting cases can be treated approximately to illustrate general trends. As shown in Equation \ref{rayleigh_rcs}, the power reflected from a single sphere with a diameter $D$ is proportional to $D^6$, when $D$ is much smaller than the radar wavelength. The particle mass itself is $\rho (4/3) \pi(D/2)^3$, where $\rho$ is the density of the material. This means that for clouds of small particles, the reflected power per cloud mass is proportional to $D^6 / D^3 = D^3$. In the large-diameter limit, where the particle size is much larger than the radar wavelength, the reflected power is simply proportional to the cross sectional area of the particle, $\pi (D/2)^2$. This means that in the large diameter limit, the reflected power per mass is proportional to $D^2 / D^3 = 1/D$. Considering these two limits means that fixing the radar operating wavelength, for small diameters the reflected power per mass should rise as $D^3$, there should be a maximum when the particle size is roughly a wavelength, and for larger sizes the power should fall as $1/D$. The limiting power laws differ slightly in our Mie scattering calculation because it includes resonance effects, and because it includes the ash particle size distribution (which somewhat averages down the resonance effects). Still, these rough scaling arguments do broadly reproduce our Mie scattering radar cross section calculations in Figures \ref{rcs_particle_size} and \ref{rcs_retuning}.

The radar operating frequency of 220 GHz was chosen because, as shown in Figure \ref{rcs_particle_size}, the reflected signal per mass is maximized for a 0.5 mm typical particle size. We chose to optimize for this size based on previous measurements of ash particle size distributions \cite{vaneaton15, gudmundsson12}. Also, since we are operating near this maximum, even a relatively large uncertainty in the size distribution of the ash particles leads to a relatively small uncertainty in the conversion from measured radar signal into ash mass density. For example, if the typical particle size $D$ were uncertain to the 1.0-0.5 mm range (i.e. $\pm33\%$ uncertainty), the uncertainty in inferred ash mass is only $\pm13\%$. This compares very favorably with conventional weather radar operating in the long-wavelength limit (i.e., signal per mass $\sim D^3$), for which the uncertainty would be large, $(-70\%,+135\%)$.

\subsubsection{In-Situ Particle Size Measurement}
\label{particle_size_text}

Even though exact knowledge of the ash particle size distribution is relatively unimportant for converting measured millimeter-wave radar signals to ash mass (Section~\ref{ash_mass}), a direct measurement of the particle size could still improve the precision of the conversion from measured radar signal to inferred ash concentration. In addition, observing the ash particle size in-situ would be an important input to validate ash transport codes, and would validate existing particle size measurement methods that rely on collecting ash from the ground after it has drifted and settled. Existing long wavelength weather radar cannot directly measure ash particle size. Harris and Rose \cite{harris83} discuss a method using existing weather radar to observe how the radar signals change through the eruption cloud, and along with assumptions about the terminal velocity of the particles they arrive at an indirect estimate of the particle size. However, since the ash particles are far smaller than its operating wavelength, varying the weather radar frequency will always follow the same $\lambda^{-4}$ scaling of Equation \ref{rayleigh_rcs}, independent of the particle size. Optical/IR measurements are also insensitive to the particle size, even by varying the observing wavelength. In the IR this is because the signal is determined by the ash temperature. At optical wavelengths, the light reflected from the ash is determined by the ash color and total cross sectional area of the cloud contents. Thus, the optical/IR signals do not indicate the individual particle size.

\begin{figure}
\begin{center}
\includegraphics[width=0.43\textwidth]{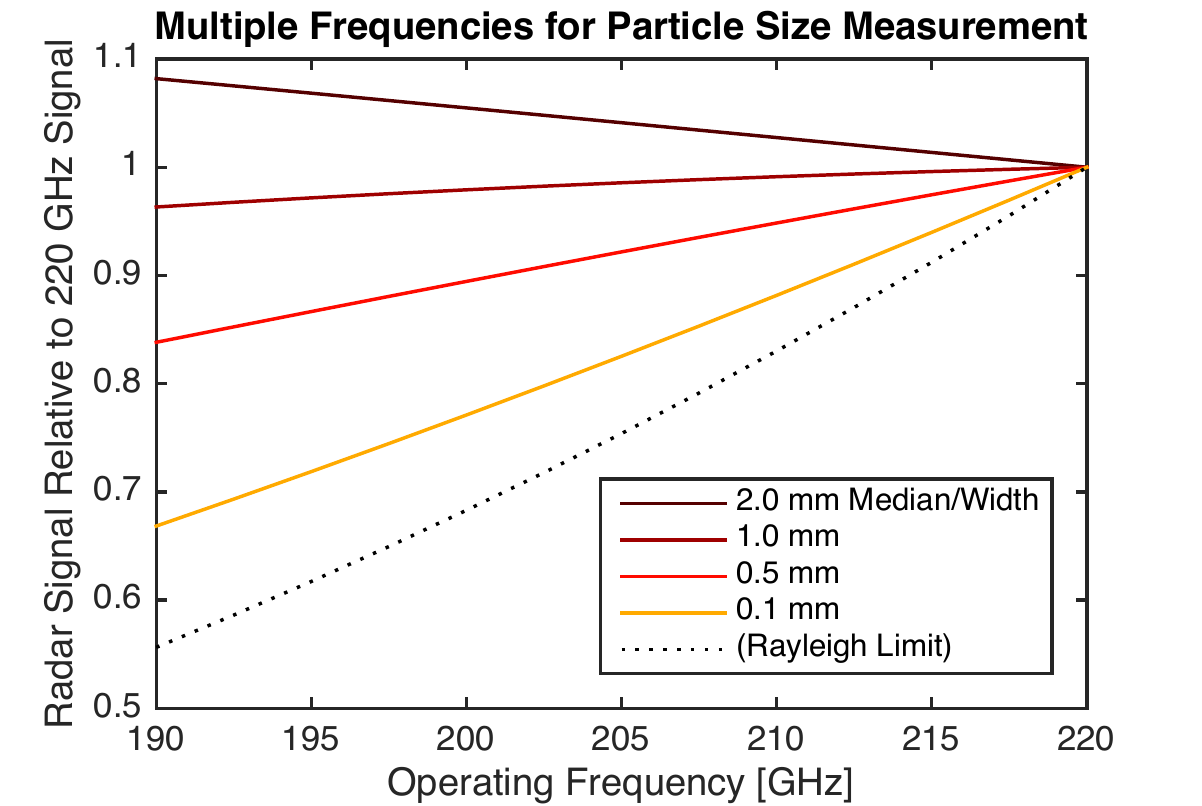}
\caption{Mie scattering calculations showing that in-situ ash particle size measurements are possible by retuning the WAMS radar in real time over the range of possible radar frequencies. (The radar frequency in WAMS is settable by changing the frequency of the local oscillator.) Since these wavelengths are similar to the ash particle size, measuring the reflected radar signal power at 220 GHz, and measuring the signal power again several other frequencies, will yield a line with a characteristic slope dependent upon particle size. A signal-to-noise of 10 or more, at several radar frequencies, would easily distinguish the different particle size distributions shown in Figure~\ref{rcs_particle_size}. Alternately, for particles much smaller than 0.1 mm, the measurement would follow the Rayleigh Limit (black dashed) line, and the data would be interpreted as a measured upper limit on particle size. Figure \ref{radar_iceland} shows that WAMS could easily achieve the required signal-to-noise with its off-the-shelf components. \label{rcs_retuning}}
\end{center}
\end{figure}

Target particle sizes can be measured by making repeated radar measurements at several wavelengths that are similar to the particle size. The application to volcanic ash is discussed briefly by Speirs and Robertson \cite{spiers11}, and here we present a detailed calculation in the context of WAMS instrument development. Since the WAMS radar wavelength is similar to the ash particle size, varying the operating frequency follows neither the $\lambda^{-4}$ scaling for Rayleigh scattering, nor the constant scaling for optical scattering. As calculated in Figure \ref{rcs_retuning} using the Mie scattering code, decreasing the radar frequency decreases the returned signal for small ash particles, stays nearly constant for particles similar in size to the wavelength, and the signal increases for large particles. This means that by retuning the WAMS radar in realtime, we can use the results of this modeling to interpret those radar measurements to yield a direct measurement of the particle size.

\section{Atmospheric Attenuation in Millimeter-Wave Remote Sensing}

\begin{table*}[]
\centering
\caption{Possible volcano observing sites, typical weather conditions, atmospheric attenuation, and required radar standoff distance.}
\footnotesize
\begin{tabular}{r|c|c|c|c|}
                        &          &             &             & \textit{Range [km] to} \\
                        &          &             & \textit{220 GHz}     & \textit{Detect 10 g/m$^3$}    \\
                        & \textit{Typical}  & \textit{Typical}     & \textit{Attenuation} & \textit{in 1 second}     \\
                        & \textit{Humidity} & \textit{Temperature} & [dB/km] & \textit{with WAMS}       \\
\textit{Location}                & [\%] & [F]     & \textit{(median)}    & \textit{(median)}          \\ \hline
Sakurajima, Japan       & 48-98    & 35-86       & \phantom{1}3.11        & 10.7              \\ \hline
Redoubt, Alaska, USA    & 41-92    & \phantom{0}2-68       & \phantom{1}1.22        & 24.3              \\ \hline
Iceland                 & 59-92    & 27-57       & \phantom{1}1.77        & 17.6              \\ \hline
Sinabung, Indonesia     & 41-99    & 69-92       & \phantom{1}5.57        & \phantom{1}6.4               \\ \hline
Stromboli, Italy        & 46-99    & 29-86       & \phantom{1}2.80        & 11.7              \\ \hline
Yasur, Vanuatu           & 46-95    & 58-89       & \phantom{1}4.51        & \phantom{1}7.7               \\ \hline             \\ \hline
\textit{(Alaska, Best-case)}     & 41       & 2           & \phantom{1}0.26        & 92.0              \\ \hline
\textit{(Indonesia, Worst-case)} & 99       & 92          & 12.42       & \phantom{1}3.1               \\ \hline
\end{tabular}
~\\
\label{atmosphere-table}
\end{table*}

Unlike at centimeter wavelengths used by conventional weather radars, millimeter-waves can be strongly attenuated by H$_2$O and O$_2$ in the atmosphere. WAMS uses this effect to measure the temperature of water vapor in a volcanic eruption near 183 GHz, where water vapor is strongly absorbing (if cold) or emitting (if warm). A key requirement for this measurement is that the column of air between the instrument and the eruption must not be completely opaque. The radar system will operate at 220 GHz, a local minimum in atmospheric absorption, but near enough in frequency to the 183 GHz water line to permit its measurement with the same antenna. Atmospheric attenuation will limit the maximum range for radar measurements. There is little atmospheric attenuation at 22 GHz.

To estimate the maximum range for millimeter-wave measurements, we calculated the atmospheric attenuation with \textit{am} \cite{paine14} for typical weather conditions throughout the year at possible observing sites. Here, unlike the full multilayer atmosphere model of Section~\ref{wams_radiometer}, we conservatively assumed that the line of sight always stayed in the full pressure and humidity of the surface atmosphere. (This is conservative because the lower pressure/humidity of the higher atmosphere has less attenuation.) There are many volcanoes worldwide that have regular eruption activity and observing locations that are accessible by 4WD pick-up. Table \ref{atmosphere-table} is a non-exhaustive list of several possible sites, including Arctic, Mid-Latitude, and Tropical sites. All of the sites we considered have low enough atmospheric attenuation to allow a typical volcanic eruption to be successfully measured under typical local weather from several kilometers away. The Alaska and Iceland sites, both of strong interest for scientific and aviation safety reasons, could be observed with the WAMS radar in typical weather from 24.3 km and 17.6 km respectively. We leave detailed planning of exact observing locations at specific sites for future work, but this initial investigation shows that good millimeter-wave measurements should be possible under a wide range of weather conditions and locations.

\section{Conclusions}

Millimeter-wave radiometer and radar systems are capable of acquiring data on ash and water vapor distributions in the interior of volcanic eruptions that are not accessible with any existing measurement technique. The 22 GHz, 183 GHz, and 232.5 GHz water vapor lines can be simultaneously monitored to measure the water vapor temperature and density profiles. Millimeter-wave radar near 220 GHz is capable of measuring fine ash clouds and directly measuring the ash particle size. Off-the-shelf components have the required sensitivity to make these measurements, which motivated our development of the WAMS instrument to make these measurements. The data will improve fundamental understanding of volcanic eruptions, and will improve realtime forecasting of the impact of fine ash clouds on aviation safety.

\bibliographystyle{unsrt}
\bibliography{bibliography}

\begin{IEEEbiography}[{\includegraphics[width=1in,height=1.25in,clip,keepaspectratio]{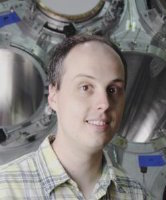}}]{Sean Bryan}
received the PhD degree from Case Western Reserve University, and is currently a postdoctoral researcher at Arizona State University. For his PhD, he worked on the Spider telescope array to measure the Cosmic Microwave Background, which flew successfully on a high-altitude balloon flight from Antarctica in the 2014-2015 season. At Arizona State, he is developing Kinetic Inductance Detectors and other millimeter-wave devices for astronomy and remote sensing applications.
\end{IEEEbiography}

\begin{IEEEbiography}[{\includegraphics[width=1in,height=1.25in,clip,keepaspectratio]{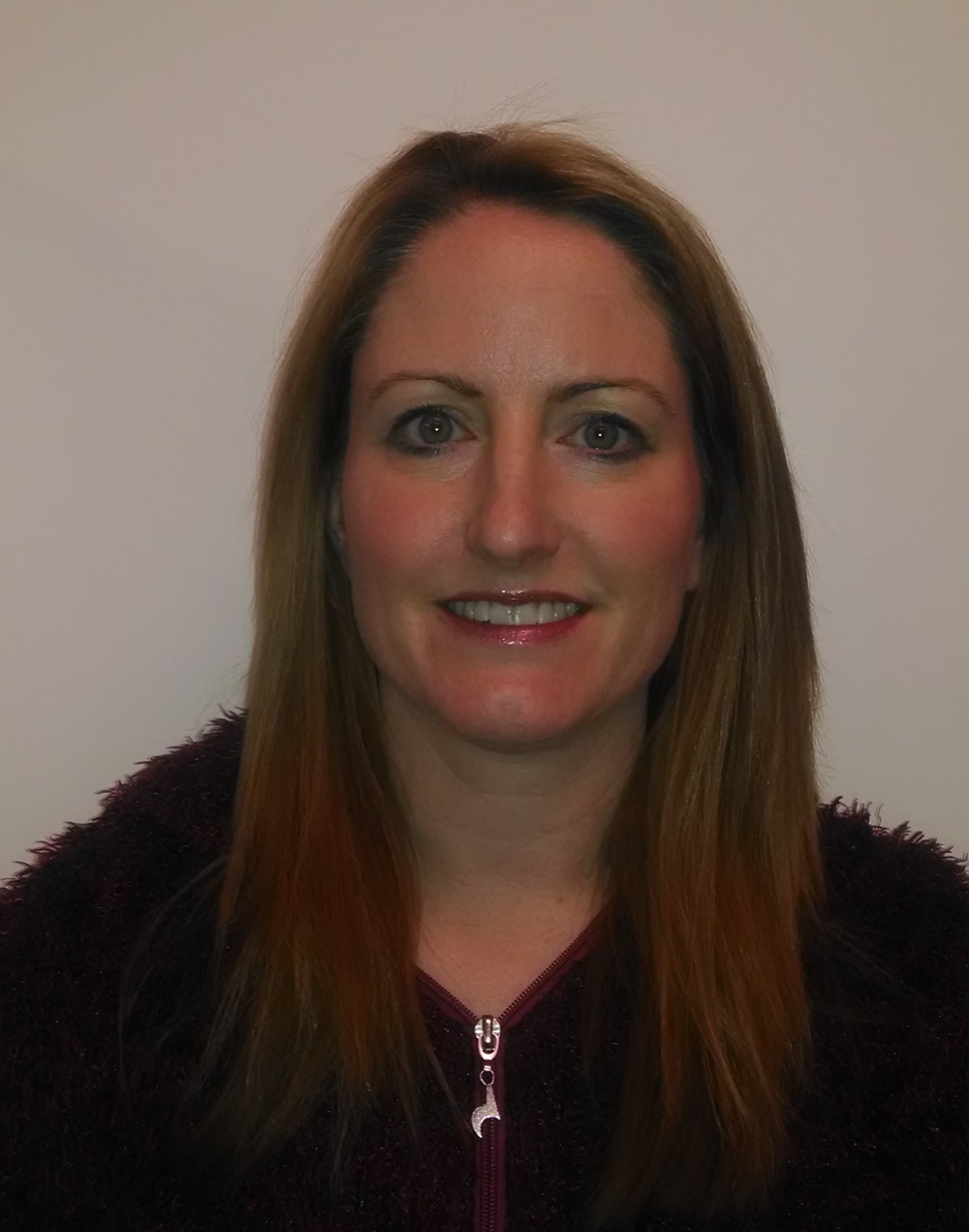}}]{Amanda Clarke} received the PhD degree in geosciences from Pennsylvania State University. She is an associate professor at the School of Earth and Space Exploration at Arizona State University. Her research interests include the physics of volcanic eruptions, developing and using numerical models and laboratory experiments to understand controls on the style and scale of volcanic eruptions, and field and satellite observation of plumes and domes to understand physical processes.
\end{IEEEbiography}

\begin{IEEEbiography}[{\includegraphics[width=1in,height=1.25in,clip,keepaspectratio]{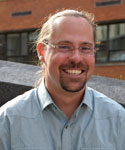}}]{Lo{\"y}c Vanderkluysen}
received the PhD degree in geology and geophysics from the University of Hawai{\char39}i, and is an assistant professor of Biodiversity, Earth and
Environmental Science at Drexel University. He investigates the cyclicity of volcanic eruptions, volcanic degassing processes, and large igneous provinces. He uses a wide array of methods, ranging from volcano monitoring and thermal remote sensing, to high-temperature geochemistry, igneous petrology, and experimental volcanology.\end{IEEEbiography}

\begin{IEEEbiography}[{\includegraphics[width=1in,height=1.25in,clip,keepaspectratio]{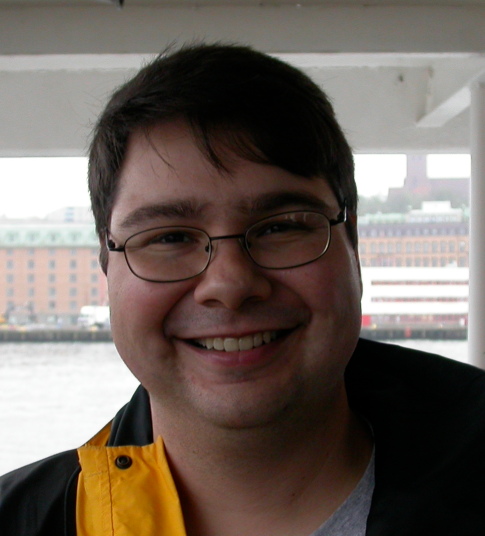}}]{Christopher Groppi}
received the PhD degree from the University of Arizona, and is an associate professor at Arizona State University. He is an experimental astrophysicist interested in the process of star and planet formation and the evolution and structure of the interstellar medium. His current research focuses on the design and construction of state of the art terahertz receiver systems to map molecular clouds. He also applies terahertz technology developed for astrophysics to a wide range of other applications including Earth and planetary science remote sensing, hazardous materials detection and applied physics.\end{IEEEbiography} 

\begin{IEEEbiography}[{\includegraphics[width=1in,height=1.25in,clip,keepaspectratio]{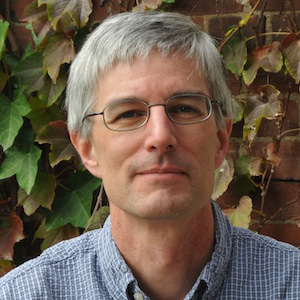}}]{Scott Paine} received the PhD degree in physics from the Massachusetts Institute of Technology. Since then, he has been with the Smithsonian Astrophysical Observatory, where he is affiliated with the Submillimeter Array. His
work includes instrumentation and techniques for
submillimeter radio astronomy and atmospheric radiometry, Fourier transform spectroscopy, and atmospheric spectroscopy and radiative transfer modeling.
\end{IEEEbiography}

\begin{IEEEbiography}[{\includegraphics[width=1in,height=1.25in,clip,keepaspectratio]{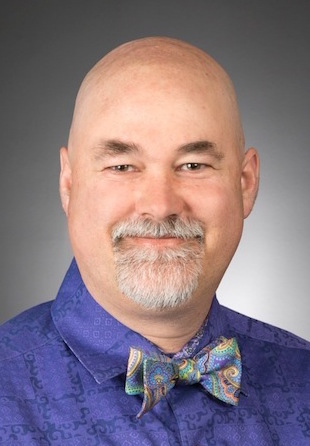}}]{Daniel Bliss} (F' 15) is an Associate Professor in the School of Electrical, Computer and Energy Engineering at Arizona State University, and the Director of ASU’s Center for Wireless Information Systems and Computational Architectures.  Dan received his Ph.D. and M.S. in Physics from the University of California at San Diego (1997 and 1995), and his B.S.E.E. in Electrical Engineering from Arizona State University (1989).  Dan was a senior member of the technical staff at MIT Lincoln Laboratory (1997-2012) in the Advanced Sensor Techniques group.  His current research topics include cooperative sensing and communications, multiple-input multiple-output (MIMO) wireless communications, MIMO radar, advanced cognitive radios, channel phenomenology, and statistical signal processing and machine learning for anticipatory physiological analysis.  Dan is a member of the IEEE AES Radar Systems Panel.
\end{IEEEbiography}

\begin{IEEEbiography}[{\includegraphics[width=1in,height=1.25in,clip,keepaspectratio]{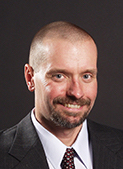}}]{James Aberle}
received the PhD degree in electrical engineering from the University of Massachusetts. As a graduate research assistant at the University of Massachusetts, Aberle developed and validated computer models for printed antennas. He is currently an associate professor of electrical engineering at Arizona State University. His research interests include the design of radio frequency systems for wireless applications as well as the modeling of complex electromagnetic phenomena.
\end{IEEEbiography}

\begin{IEEEbiography}[{\includegraphics[width=1in,height=1.25in,clip,keepaspectratio]{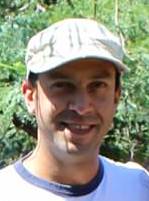}}]{Philip Mauskopf}
received the PhD degree from the University of California, Berkeley, and is a professor at Arizona State University. His background is primarily in experimental cosmology - in particular designing and building new types of instruments for measuring signals from the most distant objects in the universe. His other interests include solid state physics, atmospheric science and quantum communications and cryoptography. Before starting at ASU in 2012, he was a Professor of Experimental Astrophysics at Cardiff University in the UK since 2000 where he helped to start a world-leading group in the area of astronomical instrumentation for terahertz frequencies.\end{IEEEbiography}

\end{document}